\documentclass[12pt]{article}
\usepackage[english]{babel}
\usepackage{graphicx,epsfig}
\makeindex \textwidth 170mm \textheight 220mm \topmargin -5mm
\newcommand{\be}{\begin{equation}}
\newcommand{\ee}{\end{equation}}
\newcommand{\bea}{\begin{eqnarray}}
\newcommand{\eea}{\end{eqnarray}}

\def\bb{\bibitem}
\def\eqi{\begin{equation}}
\def\eqf{\end{equation}}
\def\eqia{\begin{eqnarray}}
\def\eqfa{\end{eqnarray}}
\def\btab{\begin{tabular}}
\def\etab{\end{tabular}}
\def\bar{\begin{array}}
\def\ear{\end{array}}

\def\GR{General Relativity}
\def\grl{general relativistic}
\def\wfs{weak--field and slow--motion approximation}
\def\leti{Lense--Thirring}
\def\grc{gravitomagnetic}

\def\se{systematic error}

\def\zh{even zonal harmonics}

\def\gp{geopotential}

\def\lg{{\rm LAGEOS}}
\def\lgg{{\rm LAGEOS} II}

\def\lb#1{\label{#1}}
\def\pc{precession}
\def\nd{node}
\def\pg{perigee}
\def\nl{nodal}

\def\sa{semimajor axis}
\def\ec{eccentricity}
\def\ic{inclination}

\def\et{Earth}
\def\ef{effect}
\def\dt#1{\dot{#1}}
\def\mlt{{\rm \mu_{LT}}}

\def\st{satellite}
\def\lt{_{\rm{LT}}}

\begin{document}
\begin{titlepage}
\begin{flushright}
\today\\
BARI-TH/00\\
\end{flushright}
\vspace{.5cm}
\begin{center}
{\LARGE A critical approach to the concept of a polar,
low--altitude LARES satellite} \vspace{1.0cm}
\quad\\
{Lorenzo Iorio$^{\dag}$\\ \vspace{0.1cm}
\quad\\
{\dag}Dipartimento di Fisica dell' Universit{\`{a}} di Bari, via
Amendola 173, 70126, Bari, Italy\\
\vspace{0.2cm}} \vspace{1.0cm}

{\bf Abstract\\}
\end{center}

{\noindent \small  According to very recent developments of the
LARES mission, which would be devoted to the measurement of the
general relativistic  Lense--Thirring effect in the gravitational
field of the Earth with Satellite Laser Ranging, it seems that the
LARES satellite might be finally launched in a polar,
low--altitude orbit by means of a relatively low--cost rocket. The
observable would be the node only. The Lense--Thirring effect on
it would consist of a secular linear trend. The biasing classical
secular nodal precessions due to the even zonal harmonics of the
geopotential, which represent the major source of uncertainty,
exactly vanished if and only if the orbit would be exactly polar.
Due to the small altitude, even small possible deviations from the
projected inclination, which might be induced by the orbital
injection errors, would yield a rather large systematic error due
to the even zonal harmonics of geopotential in the measurement of
the relativistic nodal shift. So, in this paper we show how such a
configuration, in fact, to the present level of knowledge of the
terrestrial gravitational field according to the EGM96 gravity
model, should be of relatively little utility in increasing the
obtainable accuracy in measuring the Lense--Thirring effect with
respect not only to the originally proposed supplementary
LARES--LAGEOS configuration, but also to the present
LAGEOS--LAGEOS II experiment which has a total accuracy of the
order of $20\%-30\%$. Maybe the situation might be improved, at
least to a certain extent, when the new, more accurate Earth
gravity models from the CHAMP and GRACE missions will be
available. }
\end{titlepage} \newpage \pagestyle{myheadings} \setcounter{page}{1}
\vspace{0.2cm} \baselineskip 14pt

\setcounter{footnote}{0}
\setlength{\baselineskip}{1.5\baselineskip}
\renewcommand{\theequation}{\mbox{$\arabic{equation}$}}
\noindent

\section{Introduction}
In its \wfs\ \GR\ predicts that, among other things,  the orbit of
a test particle freely falling in the gravitational field of a
central rotating body is affected by the so called \grc\ dragging
of the inertial frames or
\leti\ \ef. More precisely,  the longitude of the ascending \nd\ $\Omega$ and
the argument of the \pg\ $\omega$ of the orbit  undergo tiny \pc s
according to [\textit{Lense and Thirring}, 1918] \eqia \dot\Omega
\lt & = &
\frac{2GJ}{c^{2}a^{3}(1-e^{2})^{\frac{3}{2}}},\\
\dot\omega \lt & = &
-\frac{6GJ\cos{i}}{c^{2}a^{3}(1-e^{2})^{\frac{3}{2}}},\eqfa in
which $G$ is the Newtonian gravitational constant, $J$ is the
proper angular momentum of the central body, $c$ is the speed of
light $in\ vacuum$, $a,\ e$ and $i$ are the \sa, the \ec\ and the
\ic, respectively, of the orbit of the test particle. The
\leti\ precessions for the \lg\ satellites amount to \eqia
\dot\Omega\lt^{\rm LAGEOS}& = & 31\ \textrm{mas/y},\\
\dot\Omega\lt^{\rm LAGEOS\ II} & = & 31.5\ \textrm{mas/y},\\
\dot\omega\lt^{\rm LAGEOS} & = & 31.6\ \textrm{mas/y},\\
\dot\omega\lt^{\rm LAGEOS\ II} & = & -57\ \textrm{mas/y}. \eqfa

The first measurement of this \ef\ in the gravitational field of
the \et\ has been obtained by analyzing a suitable combination of
the laser-ranged data to the existing passive geodetic \st s \lg\
and \lgg\ [\textit{Ciufolini et al.,} 1998]. The observable [{\it
Ciufolini}, 1996] is a linear trend with a slope of 60.2
milliarcseconds per year (mas/y in the following) and includes the
residuals of the nodes of \lg\ and \lgg\ and the \pg\ of
\lgg\footnote{The \pg\ of \lg\ was not used because it introduces
large observational errors due to the smallness of the \lg\ \ec\
[{\it Ciufolini}, 1996] which amounts to 0.0045.}. The total
relative accuracy of the measurement of the solve-for parameter
$\mlt$, introduced in order to account for this \grl\ \ef, is of
the order of $2\times 10^{-1}$ [{\it Ciufolini et al.}, 1998].

In this kind of satellite--based space experiments the major
source of \se s is represented by the aliasing trends due to the
classical secular precessions [\textit{Kaula}, 1966] of the \nd\
and the \pg\ induced by the mismodelled \zh\ of the \gp\ $J_2,\
J_4,\ J_6,...$ Indeed, according to the present knowledge of the
\et's gravity field based on the EGM96 model [\textit{Lemoine et
al.}, 1998], they amount to a large part of the \grc\ precessions
of interest, especially for the first two even zonal harmonics. In
the performed LAGEOS--LAGEOS II Lense--Thirring experiment the
adopted observable allows for the cancellation of the static and
dynamical effects of $J_2$ and $J_4$. The remaining higher degree
even zonal harmonics affects the measurement at a $13\%$ level.

In order to achieve a few percent accuracy, in
[\textit{Ciufolini}, 1986] it was proposed to launch a passive
geodetic laser-ranged \st- the former {\rm LAGEOS} III - with the
same orbital parameters of \lg\ apart from its inclination which
should be supplementary to that of \lg.

This orbital configuration would be able to cancel out exactly the
classical \nl\ \pc s, which are proportional to $\cos i$, provided
that the observable to be adopted is the sum of the residuals of
the \nl\ \pc s of {\rm LAGEOS} III and LAGEOS \eqi
\delta\dt\Omega^{{\rm III}}+\delta\dt\Omega^{{\rm
I}}=62\mlt.\lb{lares}\eqf Later on the concept of the mission
slightly changed. The area-to-mass ratio of {\rm LAGEOS} III was
reduced in order to make less relevant the impact of the
non-gravitational perturbations  and the eccentricity  was
enhanced in order to be able to perform other \grl\ tests: the
LARES was born [\textit{Ciufolini}, 1998]. The orbital parameters
of \lg, \lgg\ and LARES are in Table 1.

\begin{table}[ht!]
\caption{Orbital parameters of \lg, \lgg\, LARES and POLARES.}
\label{para}
\begin{center}
\begin{tabular}{llllll}
\noalign{\hrule height 1.5pt} Orbital parameter & \lg & \lgg &
LARES & POLARES\\ \hline
$a$ (km) & 12,270 & 12,163 & 12,270 & 8,378\\
$e$ & 0.0045 & 0.014 & 0.04 & 0.04\\
$i$ (deg) & 110 & 52.65 & 70 & 90\\
\noalign{\hrule height 1.5pt}
\end{tabular}
\end{center}
\end{table}

Recent developments of the concept of the twin satellites in
supplementary orbits have led to the discover of new, possible
gravitomagnetic observables based on the use of the perigees as
well [{\it Iorio}, 2002a] and of unexpected connections with  the
gravitomagnetic clock effect [{\it Iorio and Lichtenegger}, 2002].

Unfortunately, at present we do not know if the LARES mission will
be approved by any space agency. Although much cheaper than other
proposed and/or approved  space--based missions, funding is the
major obstacle in implementing the LARES project. The most
expensive part is the launching segment.

Very recently the possibility of launching the LARES satellite
into an orbit with $a=8,378$ km, $i=90$ deg, $e=0.04$ and using as
observable its node has been considered [{\it Lucchesi and
Paolozzi}, 2001].  In the following we will name POLARES the LARES
satellite in such proposed polar orbit. The Lense--Thirring
secular rate of the POLARES node would amount to 96.9 mas/y. Then,
the observable would be \eqi\delta\dot\Omega^{\rm PL}=96.9\mu_{\rm
LT},\eqf i.e. a linear secular trend with a slope of 96.9 mas/y.
The choice of a so low altitude is motivated by the need of using
a cheap rocket launcher; so, it must certainly be considered as an
admirable and important further effort towards the practical
realization of the LARES project. The polar orbit would allow to
prevent the aliasing effects of the mismodelled classical secular
precessions of the node induced by the even $l=2n$ zonal $m=0$
coefficients of the multipolar expansion of the static part of the
geopotential. The non--gravitational perturbations [{\it
Lucchesi}, 2001; 2002; {\it Lucchesi and Paolozzi, } 2001] would
represent a minor problem.

In this paper we wish to critically analyze such important and
interesting evolution of the LARES concept.
\section{The POLARES}
\subsection{The node--only scenario}
Here we will show that the proposed POLARES configuration and the
use of the node only as gravitomagnetic observable would present
some drawbacks due to the gravitational static and time--varying
perturbations.

In regard to the classical secular precessions induced by the
centrifugal oblateness of the Earth, it should be considered that
the choice of the inclination would be of crucial importance. Such
orbital element is affected neither by the secular perturbations
induced by the even zonal harmonics of the geopotential [{\it
Kaula}, 1966] nor by the semisecular 18.6--year and 9.3--year
tidal perturbations [{\it Iorio}, 2001]. On the other hand, due to
possible orbital injection errors which are closely related to the
quality of the rocket launcher to be used, the POLARES inclination
would be far from being exactly 90 deg. The relatively low
altitude of POLARES would enhance the impact of even small
departures of $i_{\rm PL}$ from its projected nominal value on the
systematic error due to the classical static even zonal harmonics
of geopotential $\delta\mu_{\rm LT}^{\rm even\ zonals}$. This is
clearly shown in Figure 1 in which it has been calculated by means
of the covariance matrix of the EGM96 Earth gravity model up to
degree $l=20$.
\begin{figure}[ht!]
\begin{center}
\includegraphics*[width=13cm,height=10cm]{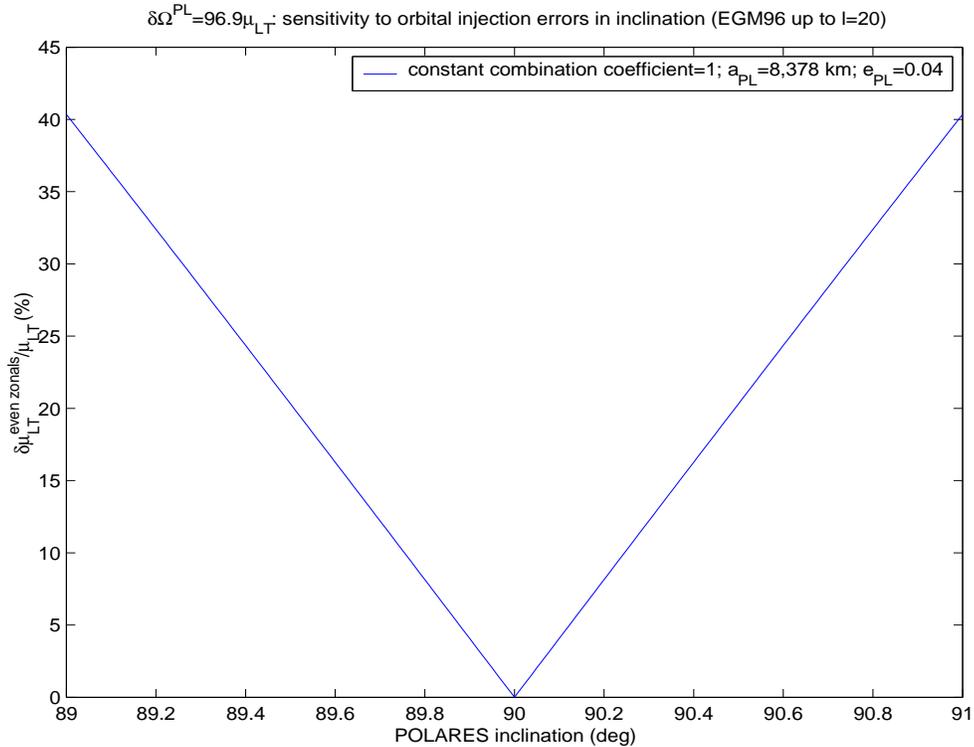}
\end{center}
\caption{\footnotesize POLARES scenario: influence of the orbital
injection errors in the POLARES inclination on the zonal error in
the Lense--Thirring nodal shift.} \label{figura1}
\end{figure}
It should be considered that, contrary to the LAGEOS satellites
which are almost insensitive to the even zonal harmonics of degree
higher than $l=20$, this would not be the case for the  POLARES,
due to its projected altitude of 2,000 km. Then, it might turn out
that the estimates of Figure 1 are optimistic.

If we consider that the impact of the mismodelled even zonal
harmonics of the geopotential on the current LAGEOS--LAGEOS II
Lense--Thirring experiment is of the order of 13$\%$, according to
EGM96, a certain weakness of the POLARES node--only scenario
becomes apparent. In regard to the possibilities offered by the
forthcoming new gravity models from the CHAMP and GRACE missions,
it should be considered that the major improvements in the
accuracy of the even zonal coefficients of the geopotential are
expected for the first degrees. Now, in the LAGEOS--LAGEOS II
Lense--Thirring experiment the first two even zonal harmonics are
cancelled out and the terms of degree higher than $l=20$ are not
relevant because the LAGEOS and LAGEOS II orbits are not sensitive
to them. So, the systematic error due to the geopotential, in this
case, is due to the even zonal harmonics from $l=6$ to $l=20$.
Then, the CHAMP and GRACE missions would certainly improve such
error. On the other hand, the POLARES configuration would be
sensitive both to the first two even zonal harmonics, which should
be greatly improved by the CHAMP and GRACE results, and to the
terms of degree higher than $l=20$, for which the improvements
should be less relevant. So, even in this perspective, maybe that
a simple reanalysis of the LAGEOS and LAGEOS II data should be
more fruitful than the node--only POLARES choice.

Such conclusion is enforced also by a simple evaluation of the
impact of some tidal perturbations\footnote{The amplitude of the
tidal perturbations on the node is proportional to $({1}/{\sin i
}){dF_{lmp}}/{di}$ where $F_{lmp}(i)$ are the inclination
functions [{\it Kaula}, 1966]. For the even zonal ($m=0$),
tesseral ($m=1$) and sectorial ($m=2$) terms they are
$F_{201}=({3}/{4})\sin^2 i-({1}/{2})$, $F_{211}=-({3}/{2})\sin
i\cos i$ and $F_{221}=({3}/{2})\sin^2 i$, respectively.} [{\it
Iorio}, 2001]. Putting aside the fact that the 18.6-year lunar
tide, which has a period of 18.6 years due to the lunar node only
motion and a large amplitude, is a $l=2,\ m=0$ constituent which
would affect the POLARES node for $i_{\rm PL}\neq 90$ deg, let us
draw our attention to the $l=2,\ m=1$ $K_1$ tesseral tide. It is
an important constituent which exerts a relevant perturbing action
on the node of a satellite\footnote{For LAGEOS it has a nominal
amplitude of the order of 10$^3$ mas [{\it Iorio}, 2001]} and has
the same period of  just the node of the satellite. It turns out
that for, say, $i_{\rm PL}=89.8$ deg the period of the POLARES
node would amount to 26,801 days, i.e. 73.3 years. This means
that, even by not considering at all the even zonal perturbations,
the mismodelled part of the $K_1$ tide itself would resemble a
superimposed aliasing trend over a reasonable observational time
span of a few years and would completely mask the Lense--Thirring
trend.
\subsection{The combined residuals scenario}
Let us try to see what could happen by inserting the node and the
perigee of POLARES in a suitable combination of orbital residuals
with the nodes of LAGEOS and LAGEOS II and the perigee of LAGEOS
II along the lines sketched in [{\it Ciufolini}, 1996; {\it
Iorio}, 2002b]. Recall that if $N$ orbital elements are present in
such combinations, the effects of the first $N-1$ even zonal
harmonics of the geopotential are cancelled out, irrespectively of
the orbital geometry of the employed satellites.

It turns out that by combining the nodes of LAGEOS, LAGEOS II and
POLARES and the perigees of LAGEOS II and POLARES in
\eqi\delta\dot\Omega^{\rm I}+c_1\delta\dot\Omega^{\rm
II}+c_2\delta\dot\Omega^{\rm PL}+c_3\delta\dot\omega^{\rm
II}+c_4\delta\dot\omega^{\rm PL}=X_{\rm LT}\mu_{\rm LT}\eqf the
inclination $i_{\rm PL}=90$ deg is singular in the sense that the
slope of the relativistic effect diverges because the coefficient
$c_2$ with which the node of POLARES enters the combination
diverges. For values of $i_{\rm PL}$ close to 90 deg Figure 2 and
Figure 3 show that the systematic error due to the remaining
$\delta J_{10}, \delta J_{12},...$, according to the covariance
matrix of EGM96 up to degree $l=20$, would seriously affect the
measurement of the Lense--Thirring linear trend. Moreover, the
impact of all the gravitational and non--gravitational
time--varying perturbations would be greatly enhanced by the large
values of $c_2$.
\begin{figure}[ht!]
\begin{center}
\includegraphics*[width=13cm,height=10cm]{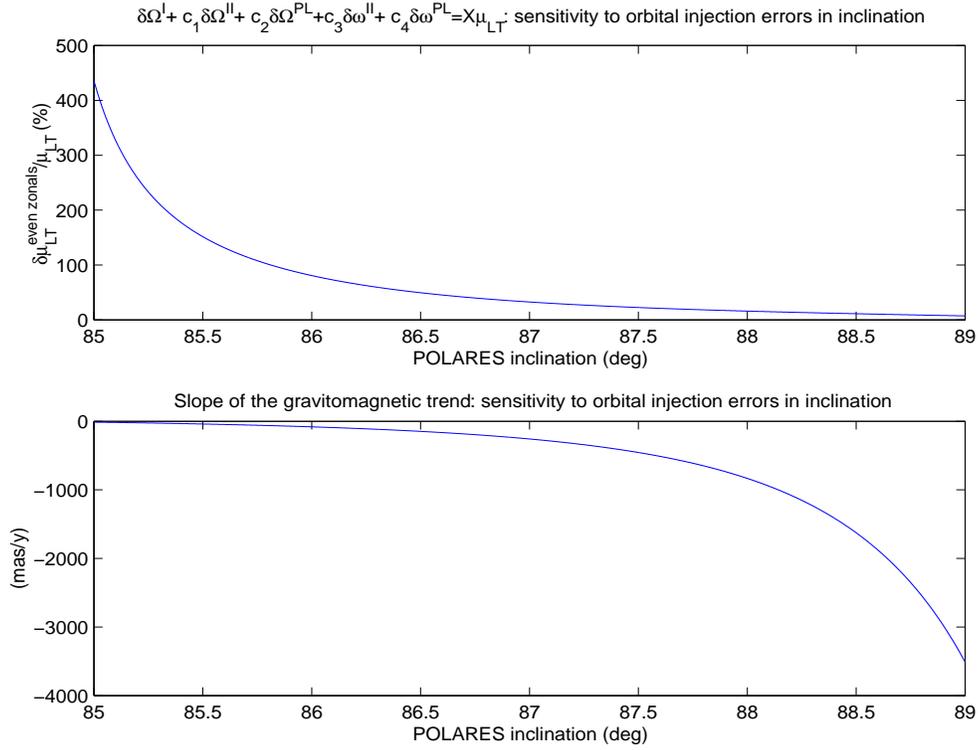}
\end{center}
\caption{\footnotesize POLARES, LAGEOS, LAGEOS II scenario:
influence of the orbital injection errors in the POLARES
inclination on the zonal error in the combined residuals with the
node of POLARES.} \label{figura2}
\end{figure}
\begin{figure}[ht!]
\begin{center}
\includegraphics*[width=13cm,height=10cm]{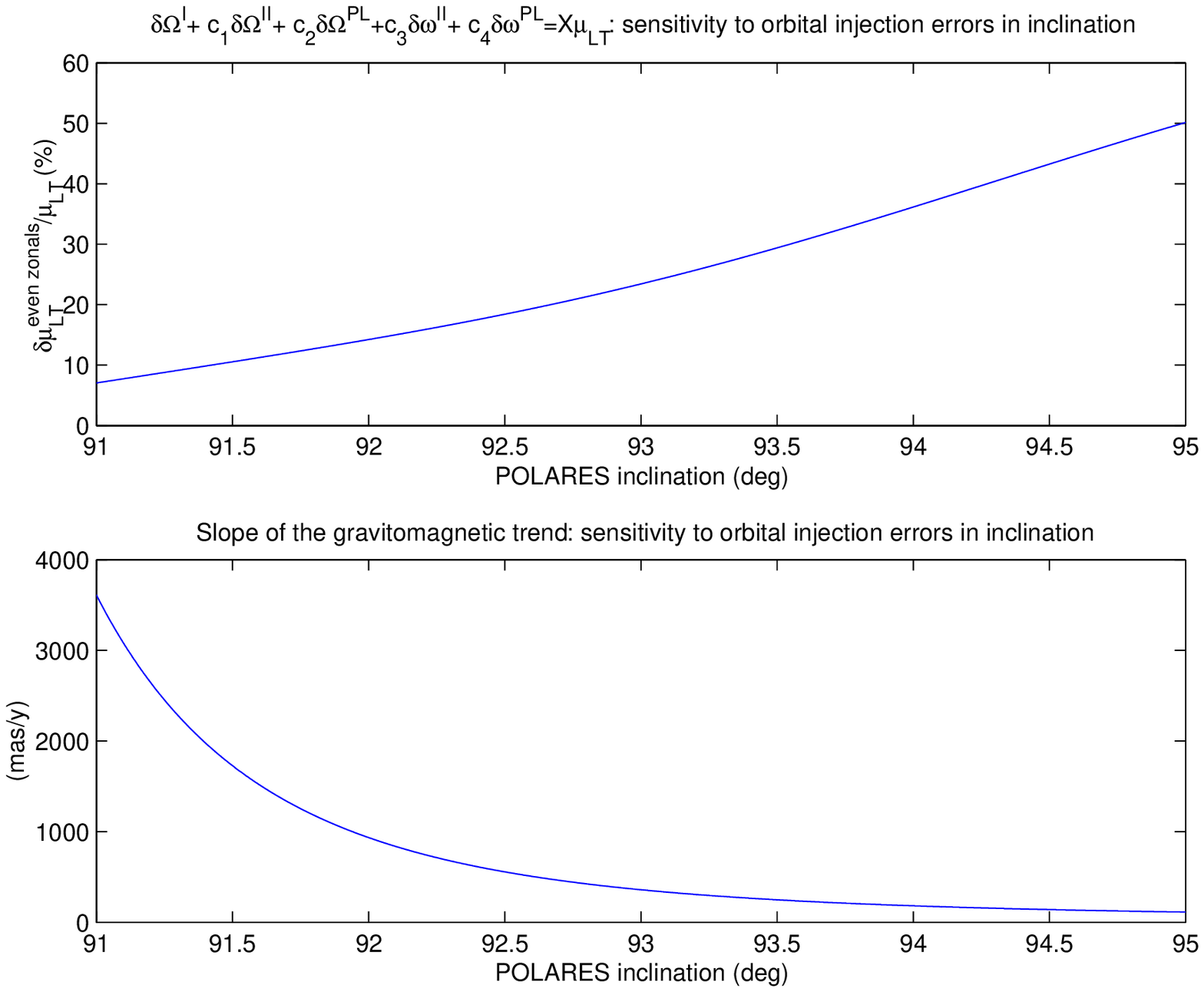}
\end{center}
\caption{\footnotesize POLARES, LAGEOS, LAGEOS II scenario:
influence of the orbital injection errors in the POLARES
inclination on the zonal error in the combined residuals with the
node of POLARES.} \label{figura3}
\end{figure}
The situation would not be better even if we would include only
the perigee of POLARES in the combined residuals
\eqi\delta\dot\Omega^{\rm I}+c_1\delta\dot\Omega^{\rm
II}+c_2\delta\dot\omega^{\rm II}+c_3\delta\dot\omega^{\rm
PL}=X_{\rm LT}\mu_{\rm LT},\eqf as shown by Figure 4.
\begin{figure}[ht!]
\begin{center}
\includegraphics*[width=13cm,height=10cm]{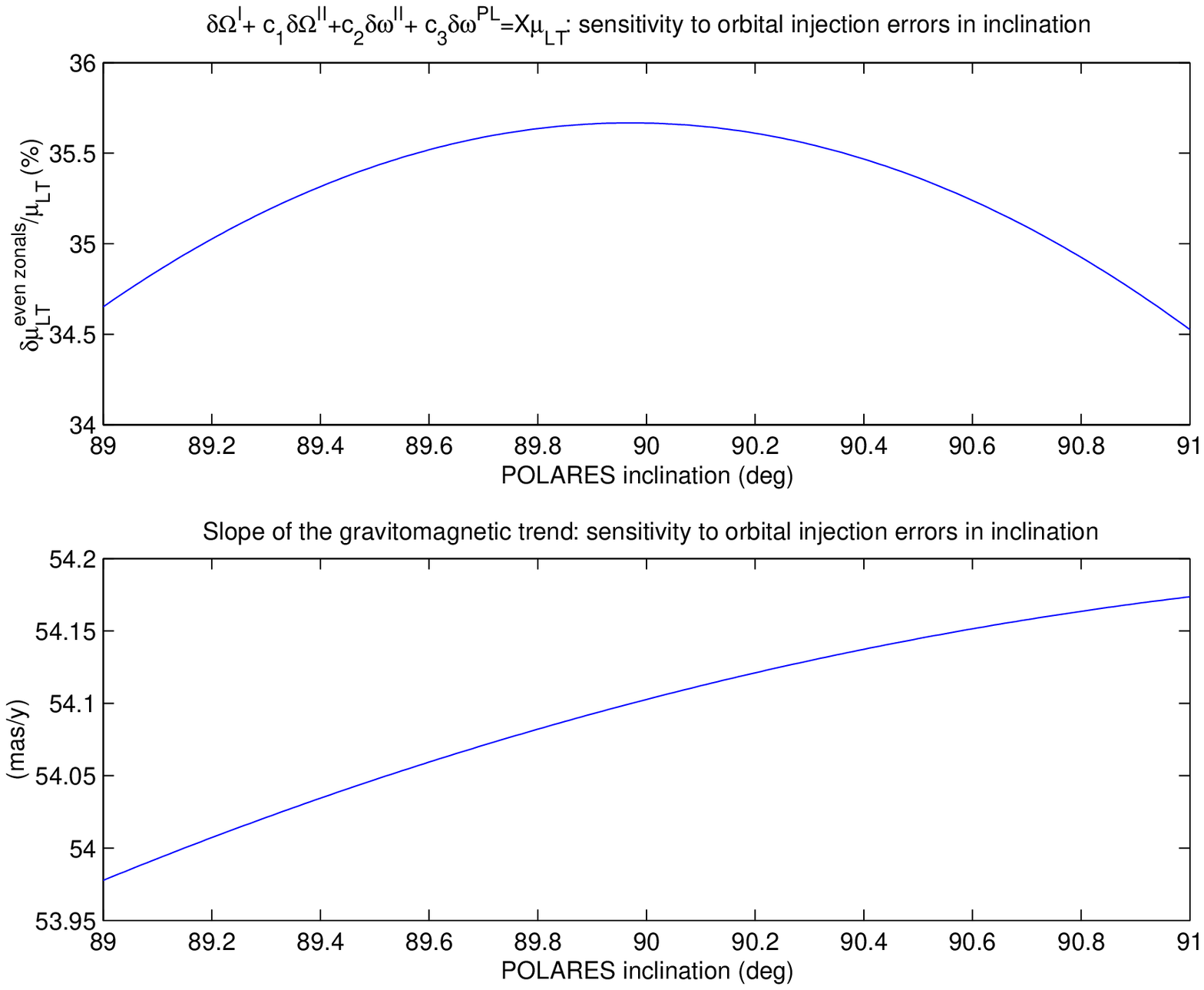}
\end{center}
\caption{\footnotesize POLARES, LAGEOS, LAGEOS II scenario:
influence of the orbital injection errors in the POLARES
inclination on the zonal error in the combined residuals without
the node of POLARES.} \label{figura4}
\end{figure}
\section{Conclusions}
In this paper the proposal of putting the LARES satellite into a
polar, elliptical orbit with an altitude of 2,000 km in order to
look at its gravitomagnetic secular node shift has been critically
analyzed.

The key point is that the mismodelled classical nodal secular
precessions induced by the even zonal coefficients of the
multipolar expansion of the terrestrial gravitational field
vanished if and only if the inclination of the satellite would be
exactly $90$ deg. Of course, mainly due to possible orbital
injection errors, this could never happen. It turns out that the
low altitude of the proposed orbital configuration, and the
consequent high sensitivity to the higher even degree zonal terms
of the geopotential, would greatly enhance the impact of even
small departures of the real values of the inclination from the
nominal value of 90 deg. For example, for just $i_{\rm PL}=90\pm
0.2$ deg the systematic gravitational error would be of the order
of 5$\%$--10$\%$, according to the EGM96 Earth gravity model up to
$l=20$, which is of the same order of magnitude of the present
LAGEOS--LAGEOS II experiment (Its total error, including various
systematic gravitational and non--gravitational perturbations, is
of the order of 20$\%$--30$\%$). Of course, such a situation would
be further made critical by the possible use of a low--cost
launcher which, unavoidably, would induce not negligible orbital
injection errors. Moreover, the tesseral $K_1$ tidal perturbation,
which has the same period of the satellite's node, would induce a
secular aliasing trend over an observational time span of a few
years because its period would amount to several tens of years for
near polar orbits.

Another important point is that the proposed POLARES would not
yield substantial improvements also in the context of the combined
residuals approach which allows to cancel out the contribution of
the first even zonal coefficients of the geopotential
irrespectively of the inclination of the satellites. Indeed, it
turns out that a combination including the nodes of LAGEOS, LAGEOS
II, POLARES and the perigees of LAGEOS II and POLARES would be not
defined for $i_{\rm PL}=90$ deg because the coefficient weighing
the node of the LARES would go to infinity. For very small
deviations of $i_{\rm PL}$ from such a critical value the
systematic error induced by the remaining even zonal harmonics
would amount to 10$\%$--20$\%$. Moreover, the coefficient with
which the node of POLARES would enter the combination would be
much more larger than unity and would greatly enhance the impact
of the gravitational and non--gravitational time--dependent
perturbations. Even the use of the nodes of LAGEOS and LAGEOS II
and the perigees of LAGEOS II and POLARES, in which case $i_{\rm
PL}=90$ deg would not create problems, would not yield benefits
because the systematic gravitational error would be of the order
of 20$\%$--30$\%$.

Moreover, the accuracy of the practical data reduction  from such
version of LARES would be affected by the atmospheric drag.

Finally, we can conclude that the proposed low--cost version of
the LARES mission would not yield any significant improvements in
the measurement of the elusive Lense--Thirring effect, according
to the present--day level of knowledge of the Earth's
gravitational field. Perhaps, the situation could improve to a
certain extent with the new, more accurate gravity models from
CHAMP and GRACE missions which should become available in the next
few years. On the contrary, the original concept of the couple of
supplementary satellites would deserve grater attention thanks to
its much richer spectrum of high accuracy relativistic
observables.

\end{document}